\documentclass[journal=jacsat,manuscript=article]{achemso}

\usepackage{graphicx}
\usepackage{dcolumn}
\usepackage{bm}
\usepackage{physics}
\usepackage{xcolor,colortbl}
\usepackage{booktabs}

\usepackage[english]{babel}
\usepackage[%
  colorlinks=true,
  urlcolor=blue,
  linkcolor=blue,
  citecolor=blue
]{hyperref}

\title{
{Modifications of Tanabe-Sugano d$^6$ diagram induced by radical ligand field: ab initio inspection of a Fe(II)-verdazyl molecular complex}
}

\author{Pablo Roseiro}
\affiliation{Laboratoire de Chimie Quantique, Institut de Chimie,
CNRS/Université de Strasbourg, 4 rue Blaise Pascal, 67000 Strasbourg, France}

\author{Saad Yalouz}
\email{yalouzsaad@gmail.com}
\affiliation{Laboratoire de Chimie Quantique, Institut de Chimie,
CNRS/Université de Strasbourg, 4 rue Blaise Pascal, 67000 Strasbourg, France}

\author{David J. R. Brook}
\affiliation{Department of Chemistry, San José State University, One Washington Square,
San José, CA 95192, USA}

\author{Nadia Ben Amor}
\affiliation{Laboratoire de Physique et Chimie Quantiques, UMR 5626 Université Paul Sabatier, 118 route de Narbonne, 31062 Toulouse, France}

\author{Vincent Robert}
\affiliation{Laboratoire de Chimie Quantique, Institut de Chimie,
CNRS/Université de Strasbourg, 4 rue Blaise Pascal, 67000 Strasbourg, France}

\date{\today} 
\begin{document}
\begin{abstract}
Quantum entanglement between the spin states of a metal centre and radical ligands is suggested in an iron(II) [Fe(dipyvd)$_2$]$^{2+}$ compound (dipyvd = 1-isopropyl-3,5-dipyridil-6-oxoverdazyl). 
Wavefunction \textit{ab initio} (Difference Dedicated Configuration Interaction, DDCI) inspections were carried out to stress the versatility of local spin states.
We named this phenonmenon \textit{excited state spinmerism}, in reference to our previous work (see Roseiro et. al., ChemPhysChem 2022, e202200478) where we introduced the concept of spinmerism as an extension of mesomerism to spin degrees of freedom.
The construction of localized molecular orbitals allows for a reading of the wavefunctions and 
projections onto the local spin states.
The low-energy spectrum is well-depicted 
by a Heisenberg picture. A 60 cm$^{-1}$
ferromagnetic interaction is calculated between 
the radical ligands with the $S_{total} = 0$  and $1$ 
states largely dominated by a local low-spin $S_{Fe} = 0$. 
In contrast, the higher-lying $S_{total} = 2$ states 
are superpositions of  the local $S_{Fe} = 1$
(17\%, 62\%)
and
$S_{Fe} = 2$ (72\%, 21\%) spin states.
Such mixing extends the traditional
picture of a high-field $d^6$ Tanabe-Sugano diagram. 
Even in the absence of spin-orbit coupling,
the avoided crossing  between different
local spin states is triggered by the 
field generated by radical ligands.
This
puzzling scenario emerges from versatile local
spin states in compounds which extend the traditional views in molecular magnetism.

\end{abstract}

\maketitle
\section{Introduction}

The synthesis and characterization of molecule-based magnetic systems has been an intense research area for decades,
prompted by the need for information storage devices and advances in quantum technologies (see for example Refs.~\cite{gaita2019molecular,troiani2011molecular,stamp2009spin,mcadams2017molecular}).  
The motivations for targeting such complexes stem from their physical-chemical properties
ranging from long coherence time~\cite{atzori2016quantum, atzori2016room, bader2014room, graham2014influence, atzori2018structural}
to manipulation possibilities.~\cite{bayliss2020optically,carretta2021perspective,li2021manipulation,nelson2020cnot,thiele2014electrically,
hussain2018coherent}
In the presence of spin-switchable units, the local spin state of archetypical 
 iron(II) or cobalt(II) ($3d^6$ and $3d^7$, respectively) ions can be controlled using
 external fields such as light or
 temperature.
 Important structural modifications are then observed, with changes in bond distances of up to 0.2 \AA~(\textit{i.e.} 10 \%)\cite{hauser2004ligand} 
 and deep changes in  charge distribution (up to 0.5 electrons).~\cite{kepenekian2009primary} The latter are responsible for 
hysteretic behavior in materials,\cite{kenepekian2009magnetic}
and have long been underestimated whereas they are the main characteristics of valence tautomers.~\cite{tezgerevska2014valence} 
More recently, transition metal ions combined with organic ligands have been considered as possible 
targets in the development of molecular-based quantum units of information, \textit{e.g.} qubits or qudits (see for example Refs.~\cite{gaita2019molecular,troiani2011molecular,stamp2009spin,mcadams2017molecular}). Theoretical studies have also revealed the potential interest of radical ligands for the manipulation of quantum information via the entanglement of local spin degrees of freedom.~\cite{roseiro2022combining,roseiro2022Qubit}  
A prerequisite is the binding of radical ligands to paramagnetic metal centers
without losing their open-shell character.
In this respect, oxoverdazyl-based ligands have proven to fulfill such requirement,
giving rise to a wealth of magnetic coupling schemes.~\cite{oms2010beyond,brook2010strong,barclay2003verdazyl} 
Furthermore, the flexibility and well-established redox activity of organic-based compounds make 
such materials particularly interesting.
Not only can inter-unit interactions be modulated with speculated
spin-crossover behaviour\cite{rota2010toward},
but the field generated by several open-shell ligands may give rise
to unusual and puzzling scenarios. 

Recently, it has been suggested that the energy spectrum of a cobalt(II) ion coordinated to
open-shell radicals may not be readily interpreted.
At the crossroads of exchange coupled and spin-crossover systems, this
compound has questioned the traditional pictures emerging from a metal ion, either high-spin or low-spin, in the electrostatic field of neighbouring ligands.~\cite{fleming2020valence,roseiro2022combining}
Wavefunction-based calculations supported that the spin states are characterized
by combinations of various local spin states on the cobalt(II) center.
In particular, the ground state displays a structure-sensitive admixture of low-spin $S_{Co} = 1/2$ in a
dominant high-spin $S_{Co} = 3/2$ , a feature of entanglement named \textit{spinmerism}.
The mixing was further interpreted by inspecting a d$^7$ Tanabe-Sugano diagram that exhibits a doublet-quartet crossing in the intermediate ligand field regime.
 \begin{figure}
    \centering
    \includegraphics[width=8cm]{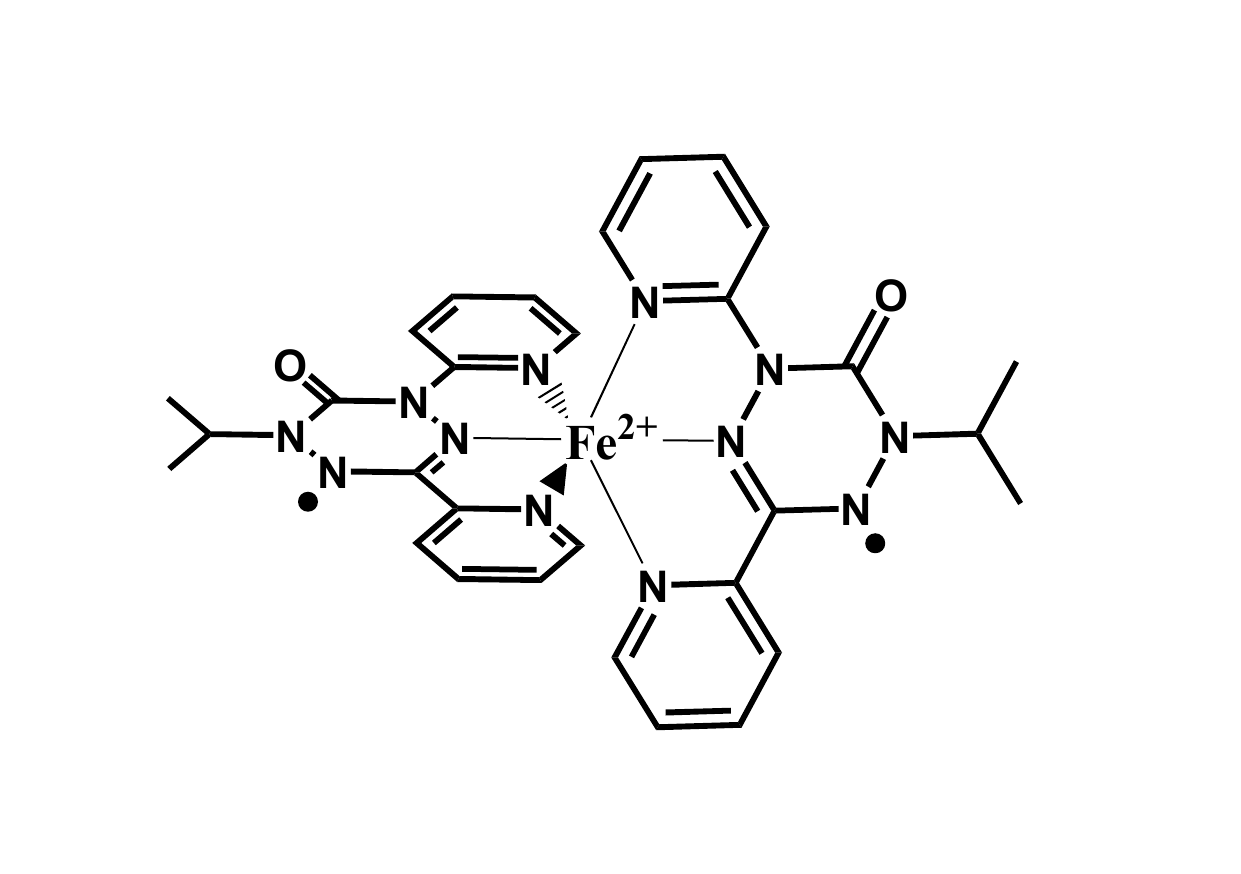}
    \caption{ \textbf{[Fe(dipyvd)$_2$]$^{2+}$ compound (1) from Ref\cite{brook2018anelectron}.}}
    \label{fig:molecule_intro}
\end{figure}
\textit{The spinmerism} phenomenon can be seen as an extension of mesomerism to spin degrees of freedom with entanglement
in between two local sub-parts of a molecule.
Since the ligand field includes Coulomb and direct exchange contributions in a complex built on spin-coupled partners, one may wonder whether different local spin states may coexist on the metal ion.
The introduction of radical ligands may indeed disrupt the assumption of a given spin state on the metal centre
in the description of ground and excited states.  

\begin{figure}
    \centering
    \includegraphics[width=8cm]{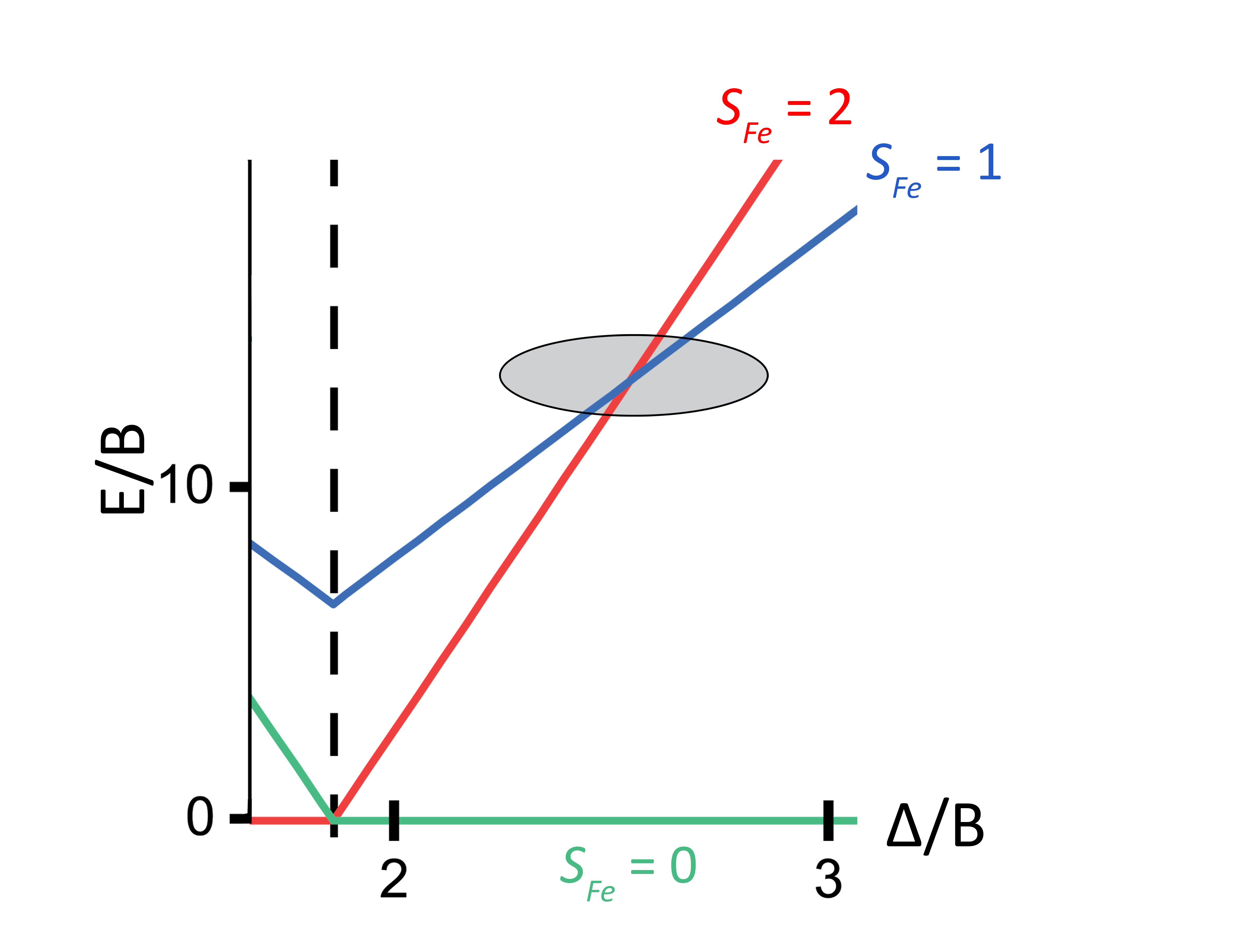}
    \caption{ \textbf{Low part of the Tanabe-Sugano diagram of an iron(II) d$^6$ ion in an octahedral field. The critical ligand field
    value marking spin-crossover is
    indicated by a vertical dotted line. The crossing between the excited $S_{Fe} = 1$ and
    $S_{Fe} = 2$  states is highlighted.}}
    \label{fig:d6_TS_diagram}
\end{figure}

Inspired by these observations, the pseudo-octahedral iron(II)-oxoverdazyl 
compound [Fe(dipyvd)$_2$]$^{2+}$, with dipyvd = 1-
isopropyl-3,5-dipyridil-6-oxoverdazyl (\textbf{1}) was considered as another prototype
to be examined (see Figure~\ref{fig:molecule_intro}).~\cite{brook2018anelectron}
Magnetic and spectroscopic properties were reported in the literature
and complemented by density functional theory-based calculations.
Iron(II) being a spin-crossover metal ion, any intermediate ligand field 
should be appropriate to observe the coexistence of a $S_{Fe} = 0$ (low spin)
and $S_{Fe} = 2$ (high spin). From the Fe-N bond distances values (\textit{ca.} 1.9 \AA),
a low-spin iron center is expected and was indeed confirmed.~\cite{brook2018anelectron}
As seen in Figure~\ref{fig:d6_TS_diagram}, the $d^6$ Tanabe-Sugano diagram exhibits 
a crossing between the excited triplet and excited quintet states. Therefore, a similar manifestation of spin states mixing observed in the ground state of cobalt(II) complex might appear in the excited states of this iron(II) analogue.
Guided by this observation in the high-field regime, we thought that
\textit{excited state spinmerism} might be anticipated, with total spin states
resulting from combinations of local $S_{Fe} = 1$ and $S_{Fe} = 2$.
For all these reasons, \textit{ab initio} calculations were carried out to inspect the
energy spectrum of \textbf{1} (shown in Fig.~\ref{fig:molecule_intro}), dominated by local spins modifications.
The examined energy window consists of spin states characterized by similar charge distributions, leaving out ligand-to-metal and metal-to-ligand charge transfer states.
The eigen-states were constructed using 
localized molecular orbitals (LMOs) to allow for projections onto the local spin states
 of the metal and ligands, $S_{Fe}$ and  $S_{L}$, respectively. 

The numerical results presented in this work highlight the presence of a strong quantum entanglement between the local spin states, \textit{i.e.} a \textit{spinmerism} effect, 
of the metal ion and the radical ligand environment. 
This evidence reshuffles the standard views in molecular magnetism and simultaneously opens up new perspectives for technological development at the interface of physics and chemistry. In Light-Induced Excited Spin-State Trapping (LIESST) experiments,~\cite{LIESST1984} the long-lived excited states at low temperatures would consist of superpositions of $S_{Fe} = 1$ and $S_{Fe} = 2$. Therefore, the variability of local spin states could provide a pathway to encode quantum information on synthetic molecular systems based on this light-induced qubit.
 
\section{Computational details}

The presence of multiple open-shells on both metal ion and ligands strongly invites the use of a wavefunction-based method,
such as the complete active space self-consistent field (CASSCF) method. The complete active space should include six electrons from the iron center and two from the oxoverdazyl ligands in seven molecular orbitals (CAS[8,7]). 
However, the active space was reduced to CAS[6,6] by inspecting the occupation numbers of the active molecular orbitals (MOs). 
The CAS[6,6]SCF sextuplet MOs were then transformed into localized molecular orbitals (LMOs)\cite{DOLOcode} which are shown in Figure~\ref{fig:MOs_cas6_6}. Such transformation allows for a reading of the wavefunctions following a Lewis-like description.
In the low-energy spectrum of \textbf{1}, we checked that the mostly metal d-type LMOs remained doubly occupied (low-spin iron(II)) and
the active space was even reduced to CAS[2,2]. 
All CASSCF calculations were performed on the available Xray structure without any geometry optimization and used the MolCAS 8.0 package.~\cite{MOLCAS} 
The iron and first coordination sphere atoms were described with 4s3p2d and 3s2p1d basis sets, respectively.
Smaller basis sets 3s2p were used for all other atoms whilst hydrogen atoms were depicted with a 1s basis set. 
The dynamical correlation and polarization effects were included following the Difference Dedicated Configuration Interaction (DDCI) method~\cite{DDCIref1,DDCIref2} as implemented in the CASDI code.~\cite{CASDIcode} Given a set of MOs, the structures of the spin states can be directly compared from the CI amplitudes.
Depending on the classes of excitations involved in the CI expansion, different levels (CAS + S, CAS + DDC2 and CAS + DDCI) can be reached beyond the CAS pictures (CAS[2,2] or CAS[6,6]). 
This variational method that follows a step-by-step construction of the wavefunction has produced a wealth of interpretations and rationalizations
with a systematic relaxation of the wavefunction (so-called "fully decontracted method").  
The resulting CI 
eigenfunctions were projected onto the local spin states of the oxoverdazyl ligands pair ($S_{L} = 0$ or $1$) 
and the iron ion ($S_{Fe} = 0$, $1$ or $2$). This procedure allows for a decomposition into the different entangled metal-ligand contributions.
To implement these projections and conduct our local spin analysis, the open access package \textit{QuantNBody}~\cite{yalouz2022quantnbody} was used.
This numerical python toolbox has been recently developed by one of us (SY) to facilitate the manipulation 
of quantum many-body operators and wavefunctions.
Based on this tool, matrix representations (in the many-electron basis) of the local metal and ligands spin operators $\hat{S}^2_{Fe}$ and $\hat{S}^2_{L}$ were built and diagonalized to access the local spin subspaces. This approach makes it possible to design spin projectors (in the many-electron basis) to target specific local spins contributions for the metal and the ligands in the multi-reference wavefunction.

\section{Results and Discussion}

\begin{figure}
    \centering
    \includegraphics[width=8cm]{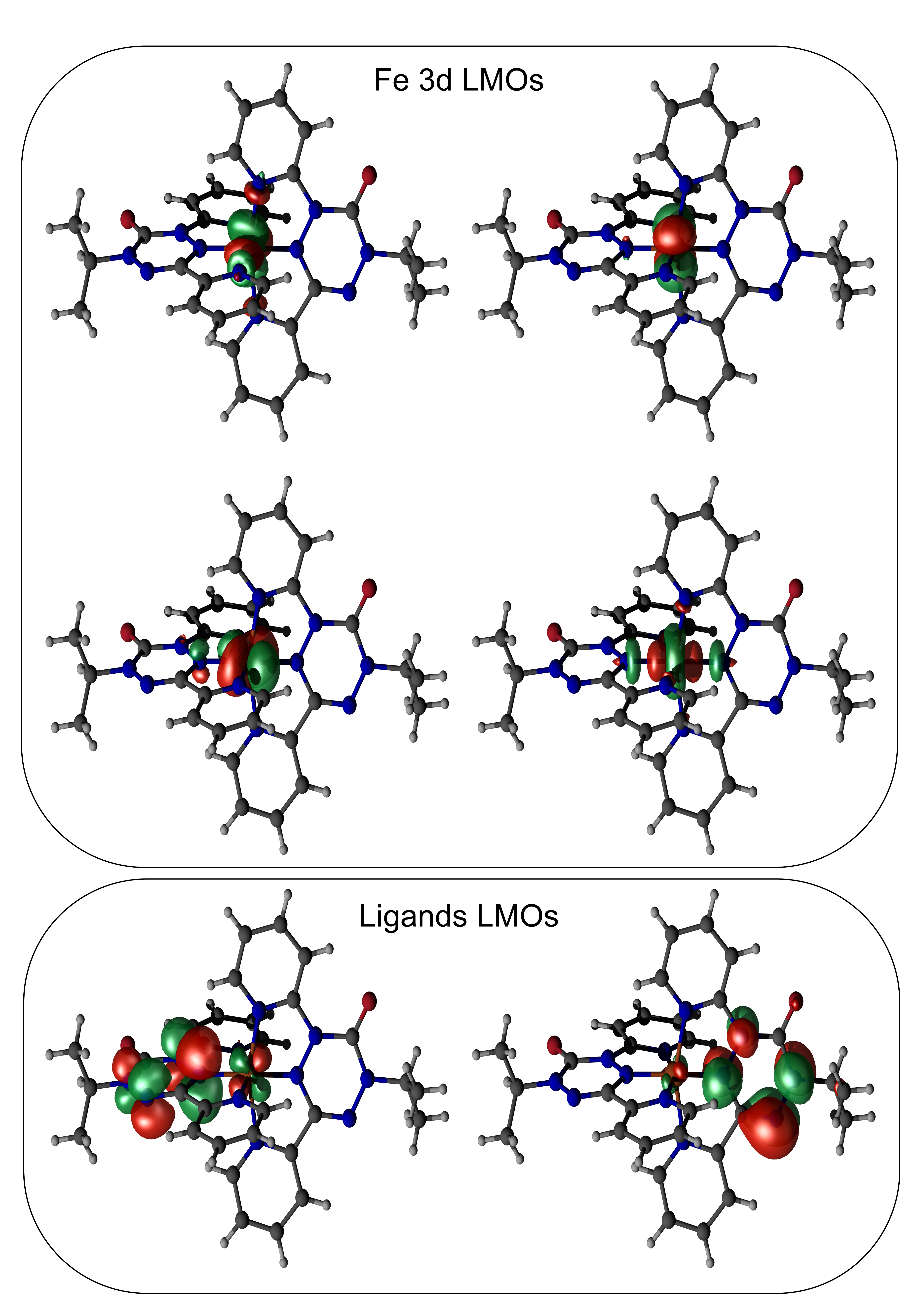}
    \caption{ \textbf{CAS[6,6] LMOs of \textbf{1} generated from a CAS[6,6]SCF calculation for the sextuplet spin state.}}
    \label{fig:MOs_cas6_6}
\end{figure}

First, CASSCF calculations were conducted on \textbf{1}. Inspections based on a 
CAS[8,7]  highlight the presence of a MO 
mostly localized on a 3d iron atomic orbital with occupation number 1.99. 
Thus, the active space was reduced down to [6,6]. The CAS[6,6]SCF MOs were localized either on the metal center  
or on each individual dipyvd ligand (see Figure~\ref{fig:MOs_cas6_6}).
Then, the electronic structure of the environment was inspected following
the procedure recently developed on a [Co(dipyvd)$_2$]$^{2+}$ compound (cobalt(II)).~\cite{fleming2020valence,roseiro2022combining}
Based on a fictitious [Zn(dipyvd)$_2$]$^{2+}$ 
(closed-shell divalent metal ion) analogue of the cobalt(II) compound, it was shown that 
the triplet-singlet energy difference is of the order of $+2$ cm$^{-1}$.~\cite{roseiro2022combining}
Considering the geometry changes moving to compound \textbf{1}, a similar inspection  was conducted by 
substituting iron(II) by zinc(II).
A 1.9 cm$^{-1}$ triplet-singlet 
energy difference was computed at the CAS[2,2] + DDCI level. 
This negligible exchange coupling value between the radical
ligands suggests that any projection of the total spin states of \textbf{1} may 
simultaneously involve the local $S_L = 0$ and $S_L=1$ states. Besides, this value can be seen as a reference to quantify the role of the electronic structure of the bridging metal ion. Indeed, the nature, and more importantly the spin state, of the metal center are likely to modify the exchange coupling constant value, a particular issue we wanted to inspect.
As reported in the literature, the intramolecular radical-radical exchange couplings were measured in a series of M(II)-(bipyvdz)$_2$ complexes (M = Mn, Ni, Cu, Zn) and range from weakly antiferromagnetic ($-10$ cm$^{-1}$) to  weakly ferromagnetic ($+2$ cm$^{-1}$).~\cite{barclay2003verdazyl}

Due to the system size and the number of open-shells, a CAS[6,6] + DDCI level of calculation is out of reach.
Thus, all our conclusions are based on CAS[6,6] + DDC2 
excitation energies and 
corresponding wavefunctions analysis, as summarized in Table~\ref{table:DDCI2_results}.
\begin{table}[h!]
\centering
\begin{tabular}{| m{2cm} |   m{2cm} |   m{2cm}  |   m{2cm}  |  m{2cm} |} 
 \hline 
  Label & Energy ($cm^{-1}$)& $2S_{total}+1$ & $S_{Fe}$ & $S_{L}$ \\  
 \hline\hline
  \rowcolor{gray!45} $Q_2$ \hfill $\bullet$& 10156 & 5 & 86\% \textit{Q} & {  42\% \textit{S}  \par  45\% \textit{T} } \\
  \hline
  \rowcolor{gray!20} $S_1$ \hfill $\star$& 8603 & 1 & { 82\% \textit{T} } & { 82\% \textit{T} } \\
  \hline
  \rowcolor{gray!20} $T_3$ \hfill $\star$& 7691 & 3 & 83\% \textit{Q} & 83\% \textit{T} \\
  \hline
  \rowcolor{gray!45} $T_2$ \hfill $\bullet$& 6705 & 3& 87\% \textit{T} & {  9\% \textit{S}  \par  79\% \textit{T} } \\
  \hline
  \rowcolor{gray!20} $H_0$ \hfill $\star$& 5326 & 7 & 100\% \textit{Q} & 100\% \textit{T} \\
  \hline
  \rowcolor{gray!45} $T_1$ \hfill $\bullet$& 4972 & 3 &     { 81\%\textit{T} }    &   {  78\% \textit{S}  \par  3\% \textit{T} } \\
  \hline
  $Q_1$ & 3323 & 5 &     { 62\%\textit{T} \par 21\% \textit{Q} }    &   {  11\% \textit{S}  \par  72\% \textit{T} } \\
  \hline
  $Q_0$ & 2906 & 5 &     { 17\% \textit{T}  \par  72\% \textit{Q}} & {  39\% \textit{S}  \par   50\% \textit{T}} \\
  \hline
  \rowcolor{gray!20} $S_0$ \hfill $\star$& 120 & 1 &     { 84\% \textit{S}} & {  84\% \textit{S}  } \\
  \hline
  \rowcolor{gray!20} $T_0$ \hfill $\star$ & 0 (ref.) & 3 & 87\% \textit{S} & 87\% \textit{T} \\
 \hline
\end{tabular}
\caption{\textbf{Energy spectrum of \textbf{1}. Quintets and  heptuplet energies are
calculated at the CAS[6,6] + DDC2 level. 
The ground state energy is used as a reference energy. Spin multiplicities $2S_{total}+1$, local spin proportions ($S_{Fe}$ and $S_L$) are given. $S$, $T$ , $Q$ and $H$ correspond to singlet, triplet, quintet and heptuplet, respectively.
}}
\label{table:DDCI2_results}
\end{table}

\begin{figure}
    \centering
    \includegraphics[width=10cm]{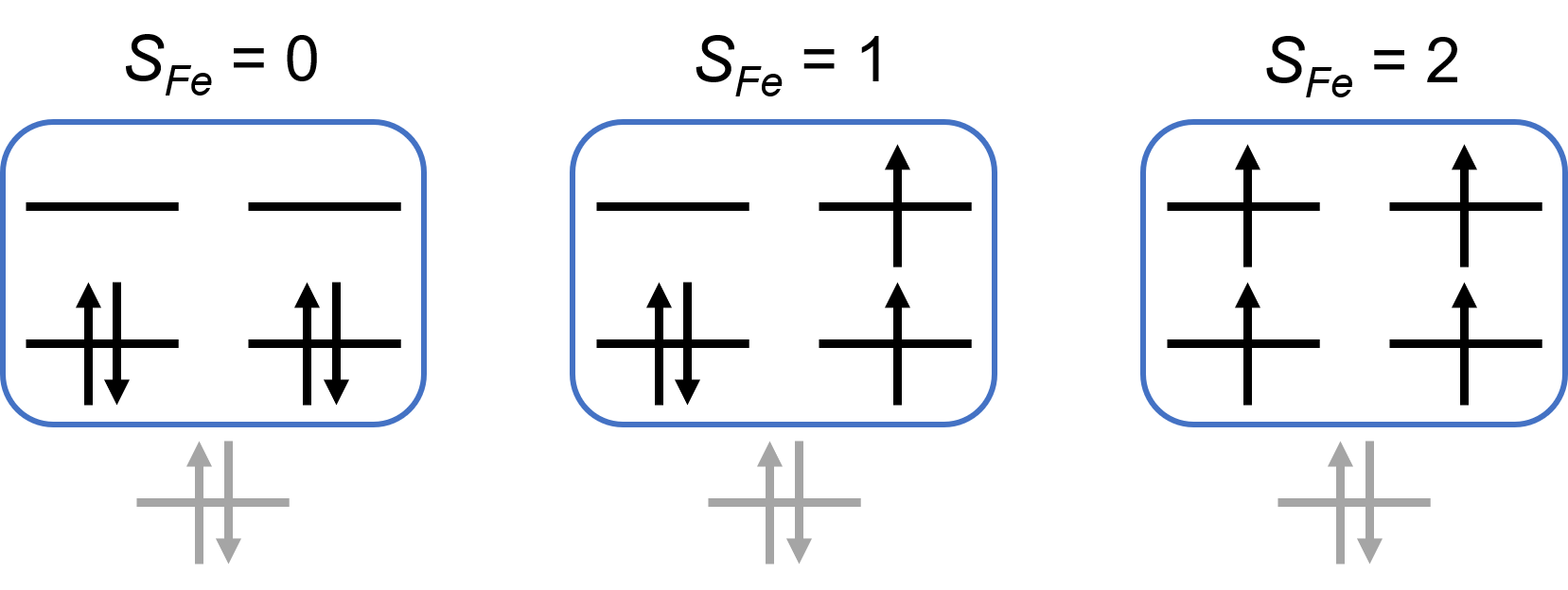}
    \caption{ \textbf{Electronic configurations within the active space associated with the local $S_{Fe}$ spin states
    arising in the energy spectrum of \textbf{1}. 
    For the sake of simplicity, the maximum spin projections
    are shown.}}
    \label{fig:open_shells}
\end{figure}

The energy spectrum results from the local spin states $S_{Fe}=0,1,2$ and $S_L=0,1$ which give rise
to ten states, namely two singlets ($S_i$, $ i=0-1$), four triplets ($T_i$, $i=0-3$), three quintets
($Q_i$, $i=0-2$) and a single heptuplet ($H_0$).
From our calculations, the local spin states display 0, 2 and 4  open-shells
on the $S_{Fe}=0$,  $S_{Fe}=1$ and  $S_{Fe}= 2$, respectively (see  Figure~\ref{fig:open_shells}).
However, the local spin values can result either from pure spin states or superpositions of different spin 
multiplicities as manifested in spinmerism.~\cite{roseiro2022combining,roseiro2022Qubit}
Therefore, four different classes can be {\it a priori} anticipated in the energy spectrum of \textbf{1}.

\begin{figure}
    \centering
    \includegraphics[width=8cm]{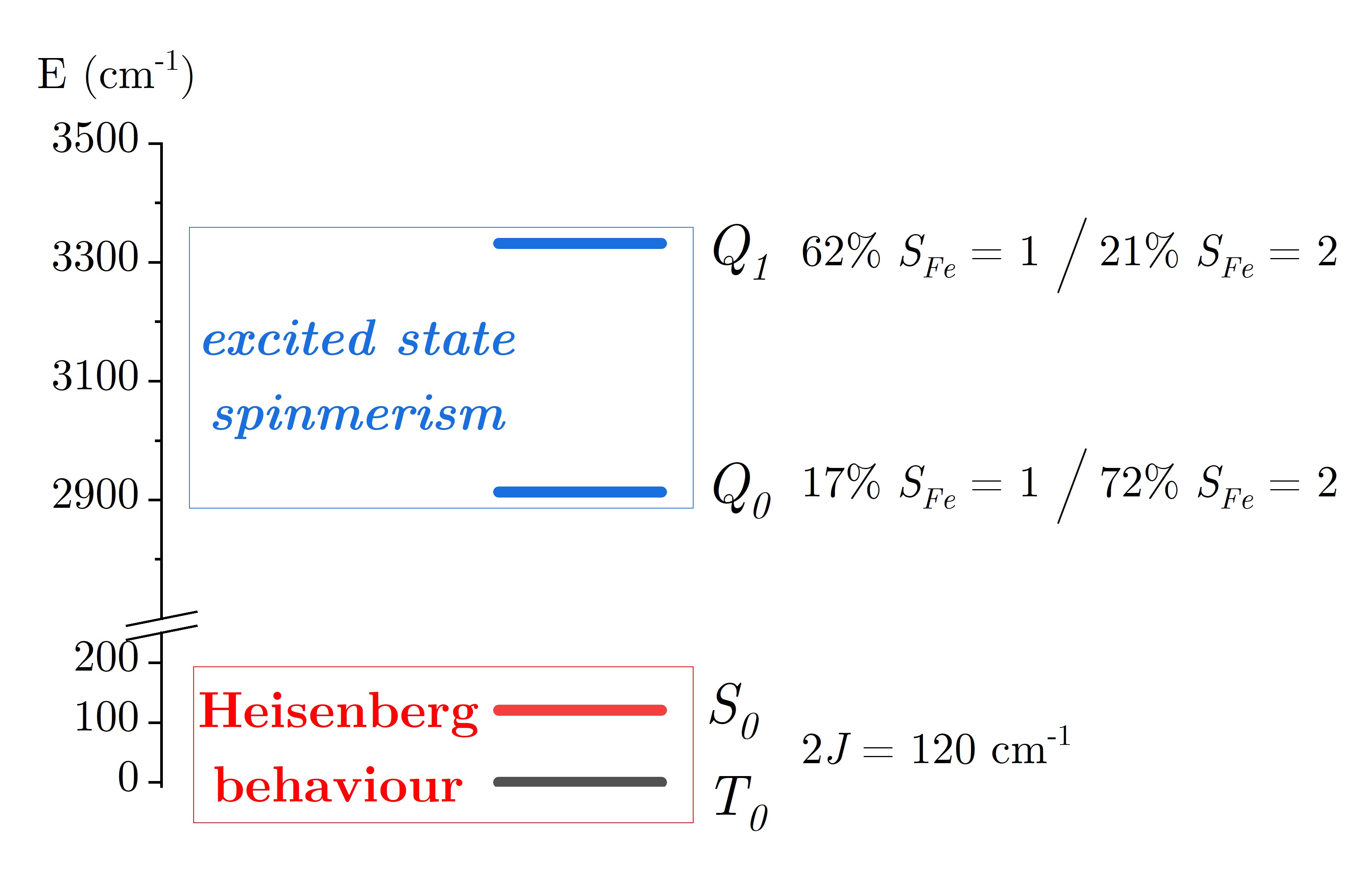}
    \caption{ \textbf{Energy spectrum of \textbf{1} reflecting a Heisenberg behaviour in the low-energy part with $T_0$ and $S_0$ states, whereas \textit{excited sate spinmerism} is observed higher in energy with $Q_0$ and $Q_1$ states. The excited quintet states $Q_0$ and $Q_1$} 
    exhibit a strong mixing between the $S_{Fe} = 1$ and $S_{Fe} = 2$ local spin states.}
    \label{fig:spectrum_fevdz}
\end{figure}

Let us concentrate on the first class of
states  dominated by pure local spin states $S_{Fe}$ and $S_L$ (see light grey entry in Table~\ref{table:DDCI2_results}).
First, the ground state $T_0$ is  triplet,  
dominated by  the local spin states  $S_{Fe} = 0$ and $S_L = 1$, respectively, followed by the singlet $S_0$  lying $120$ cm$^{-1}$ above. This picture is consistent
with a $d^6$ low-spin metal ion in a high-field environment. One should note that small admixtures ($13-16\%$) of charge transfers ($S_{Fe} = 1/2$ and $S_L=1/2$)
are evidently observed.
Since the CAS[6,6]SCF occupation numbers of the mostly d-type LMOs are larger than 1.98,
the  $S_0 - T_0$ energy difference was further
confirmed from CAS[2,2] + DDCI calculations (CAS[2,2]SCF triplet MOs) and turned out to be $120$ cm$^{-1}$. This value is in reasonable agreement
with the reported one which may vary depending on the extraction from density functional theory broken-symmetry
calculations.~\cite{brook2018anelectron}
Thus, the low-energy part of the spectrum of \textbf{1} can be viewed as
two organic radicals coupled through a closed-shell iron(II) (see Figure~\ref{fig:spectrum_fevdz}).
Evidently, a  Heisenberg Hamiltonian $\hat{H} = -2J \hat{s}_{1}  \hat{s}_{2}$
can be derived from the singlet-triplet energy difference. The
strong ferromagnetic exchange coupling constant $J = +60$ cm$^{-1}$ accounts for the low-temperature magnetic properties experimentally observed.~\cite{brook2018anelectron} This coupling is significantly larger than the reference one we estimated to be $+1.9$ cm$^{-1}$ for the closed-shell hypothetical zinc(II) complex. Such contrast is also found with the experimental values reported in a series of compounds, that not only differ from the relative positions of the verdazyl ligands but also from the absence of the iron(II) complex in the series.~\cite{barclay2003verdazyl}
The higher-lying states  $T_3$,  $S_1$ (and evidently the heptuplet $H_0$) are all characterized by 
a pure  $S_L = 1$ spin state on the coordination sphere. These states are expected from the traditional  d$^6$ Tanabe-Sugano diagram (see
Figure~\ref{fig:d6_TS_diagram}). In a triplet $S_L = 1$ high-field regime, the states ordering 
follows $S_{Fe} = 0$ ($T_0$, reference energy), $S_{Fe} = 2$ ($T_3$, 7691 cm$^{-1}$), 
and $S_{Fe} = 1$ ($S_1$, 8603 cm$^{-1}$). At this stage, the main difference with usual picture
in coordination chemistry compounds lies in the  triplet nature of the ligand field, a concept
introduced in the literature as
”excited state coordination chemistry”.~\cite{ghosh2003noninnocence}
One should finally mention the presence of 1.6\%  singlet state on the metal in $S_1$.
However, this negligible contribution arises from an excited open-shell singlet.

Based on this preliminary observation suggesting a $S_L = 1$ ligand field, we now examine the third $Q_1$ ($2 S_{total} + 1 = 5$)
excited state given in Table~\ref{table:DDCI2_results}. The iron(II) spin state is dominated by
a $S_{Fe} = 1$ (62\%) in the field of a 72\% $S_L = 1$.
Again, such picture obtained from the
spin projections of the wavefunction is consistent with a high-field $d^6$ Tanabe-Sugano diagram.~\cite{tanabe1954onthe}
However, the quintet state $Q_1$ in Figure~\ref{fig:spectrum_fevdz}
 exhibits non-negligible  contributions  from 
the local  $S_{Fe} = 2$ (21\%) and $S_{L} = 0$ (11\%) spin states. Such superposition of
triplet and quintet metal spin states in the absence of spin-orbit coupling is
a manifestation of the recently reported \textit{spinmerism} effect.~\cite{roseiro2022combining}
Our analysis supports the appearance of a similar phenomenon 
in \textbf{1}, named as
\textit{excited state spinmerism}.
This manifestation results from the open-shell
character of the environment and the energy crossing between the $S_{Fe} = 1$ and $S_{Fe} = 2$ excited states in
the high-field regime of the iron(II) $d^6$ diagram (see Figure~\ref{fig:d6_TS_diagram}). 
The traditional allowed crossing between different spin multiplicities in the Tanabe-Sugano diagram is lifted from the 
presence of two radical ligands.
As expected, the energy spectrum exhibits a second close-in-energy quintet state resulting from this mixing (see $Q_0$ in Figure~\ref{fig:spectrum_fevdz}).
The latter is found 2906 cm$^{-1}$ above the ground state, with a dominant $S_{Fe} = 2$ character (72\%), and 
a significant mixing between the environment $S_L = 0$ (39\%) and   $S_L = 1$ (50\%) spin states.
This second class of states  ($Q_0$  and  $Q_1$ in Table~\ref{table:DDCI2_results}) 
displays entanglement between the spin states of the metal ion and its coordination sphere 
in the high-field regime. In agreement with a recent inspection based on a 
model Hamiltonian~\cite{roseiro2022Qubit}, the emergence of the two quintet  states $Q_0$ and $Q_1$ in Figure~\ref{fig:spectrum_fevdz}
from avoided crossing stresses the importance of not only the
magnitude of the ligand field (Coulomb contributions), but also the
open-shell character of such field (decisive exchange contributions).

From these numerical inspections, one may further take advantage of the electronic structure changes observed in  the photo-induced spin-states of spin crossover compounds (LIESST effect).
Under low-temperature irradiation, it is possible to
quantitatively achieve a  low-spin to high-spin 
conversion in iron(II) spin-crossover compounds.~\cite{LIESST1984}
Very recently, it has been shown that unusually long
relaxation times can be reached (\textit{ca.} 20 hours)
which makes such complexes particularly encouraging
for practical applications.~\cite{delgado2018verylong}
In compound \textbf{1}, the photo-generated states would be 
$Q_0$ and $Q_1$, combinations of
the local $S_{Fe} = 1$ and $S_{Fe} = 2$.
The mechanism of light-induced spin crossover was previously studied 
to calculate intersystem-crossing rates and concluded on a process mediated by a triplet excited state.~\cite{sousa2013ultrafast,alias2022quantum}
From the  manifestation of  the here-proposed
\textit{excited state spinerism} phenomenon,
the LIESST effect would produce
local superpositions of iron(II) spin states with
potential applications as spin-qubits.

A third class of states in the energy spectrum of \textbf{1} is characterized by
a pure local spin state on the metal and a superposition of $S_{L} = 0$ and $S_{L} = 1$ on the
environment ($T_1$,  $T_2$, $Q_2$, see dark grey entry in Table~\ref{table:DDCI2_results}).
Such mixture was previously reported in noninnocent ligand-based  iron(III)
compounds~\cite{messaoudi2006correlated,guihery2008abinitio}
where the "excited state coordination
chemistry" concept~\cite{ghosh2003noninnocence}
was numerically evidenced. In compound \textbf{1}, charge transfers are small enough 
to maintain a formally iron(II) ion with spin crossover behavior, a prerequisite for spinmerism
manifestation.

Finally, our analysis of the low-lying spin states does not reflect any 
representative of the fourth class, characterized by the superposition of
$S_{Fe}$ values in the field of a pure $S_{L}$ value. Evidently, a
$S_{L} = 0$ ligand field would not allow the mixing of different $S_{Fe}$ values. However, the
absence of a mixed-spin state on the metal in a $S_{L} = 1$ ligand field is more puzzling at first.
As recently reported, the entanglement between local spin states is directly controlled by 
the relative amplitudes of the direct exchange values and conditions which might
not be fulfilled in the examined compound \textbf{1}.~\cite{roseiro2022Qubit}

\section{Conclusion}
The spin states structures of a coordination compound [Fe(dipvdz)$_2$]$^{2+}$ built on a spin-crossover 
ion (iron(II)) and two  radical ligands (oxoverdazyl) were analyzed from DDCI wavefunctions calculations. 
The procedure is based on the generation  of localized molecular
orbitals and spin projections
onto the local spin states. A ground state triplet $S_{total} = 1$ characterized by a $S_{Fe} = 0$ local spin state
is found, a reflection of a high-field regime.
A strong ferromagnetic interaction $J = +60$ cm$^{-1}$   is calculated
featuring the coupling of organic radical spin holders through a low-spin metal ion. 
Even though the low-energy part of the spectrum can be rationalized by a Heisenberg spin Hamiltonian,
the magnetic picture delivered by a d$^6$ Tanabe-Sugano diagram is deeply reshuffled in
the next nearest excited states. The flexibility afforded by the open-shell character of the environment gives rise
to marked superpositions of local spin states $S_{Fe} = 1$
and $S_{Fe} = 2$  in the $S_{total} = 2$ excited states.
The traditional quasi-degeneracy in the high-field regime is lifted by 417 cm$^{-1}$
with significant contributions from both $S_{L} = 1$ and $S_{L} = 0$ on the ligand pair.
This observation, which we name \textit{excited state spinmerism}, extends a phenomenon that was reported in a
cobalt(II) analogue.
The prerequisite for the manifestation of such entanglement is fulfilled
from the presence of an iron(II) ion and radical organic ligands.
By irradiating the sample in the UV-vis or near-IR
regions at low temperature, such compounds might be photo-switched and become original targets.
Our analysis may stimulate experimentalists to photo-generate slowly decaying excited states 
consisting of $S_{Fe} = 1$ and $S_{Fe} = 2$ spin states superpositions. Therefore, the variability of local spin states could 
provide a pathway to encode quantum information on synthetic molecular systems.
This particular class of coordination chemistry compounds combining versatile local spin states
not only enlarges the traditional pictures in molecular magnetism but 
might become original targets for spin-qubit generation.

\section*{Acknowledgements}
This work was supported by the Interdisciplinary Thematic Institute SysChem via the IdEx Unistra (ANR-10-IDEX-0002) within the program Investissement d’Avenir.
P. R. acknowledges the Ecole Doctorale de
Sciences Chimiques de Strasbourg, EDSC222, and the french minister for financial support.
D.J.R.B acknowledges the support of the National Science Foundation Grant CHE-1900491.
The authors would like to thank Pr. C. Train, Dr. G. Novitchi and Dr. S. Stoian for useful discussions.

\section*{AUTHOR DECLARATIONS}

\subsection*{CONFLICTS OF INTEREST}
The authors have no conflicts to disclose.

\subsection*{ DATA AVAILABILITY }
The data that support the findings of this study are available from the corresponding author upon reasonable request.

\appendix

\numberwithin{equation}{section}
\setcounter{equation}{0}
 
\bibliography{biblio.bib}

\newcommand{\Aa}[0]{Aa}
\providecommand{\latin}[1]{#1}
\makeatletter
\providecommand{\doi}
  {\begingroup\let\do\@makeother\dospecials
  \catcode`\{=1 \catcode`\}=2 \doi@aux}
\providecommand{\doi@aux}[1]{\endgroup\texttt{#1}}
\makeatother
\providecommand*\mcitethebibliography{\thebibliography}
\csname @ifundefined\endcsname{endmcitethebibliography}
  {\let\endmcitethebibliography\endthebibliography}{}
\begin{mcitethebibliography}{42}
\providecommand*\natexlab[1]{#1}
\providecommand*\mciteSetBstSublistMode[1]{}
\providecommand*\mciteSetBstMaxWidthForm[2]{}
\providecommand*\mciteBstWouldAddEndPuncttrue
  {\def\EndOfBibitem{\unskip.}}
\providecommand*\mciteBstWouldAddEndPunctfalse
  {\let\EndOfBibitem\relax}
\providecommand*\mciteSetBstMidEndSepPunct[3]{}
\providecommand*\mciteSetBstSublistLabelBeginEnd[3]{}
\providecommand*\EndOfBibitem{}
\mciteSetBstSublistMode{f}
\mciteSetBstMaxWidthForm{subitem}{(\alph{mcitesubitemcount})}
\mciteSetBstSublistLabelBeginEnd
  {\mcitemaxwidthsubitemform\space}
  {\relax}
  {\relax}

\bibitem[Gaita-Ari{\~n}o \latin{et~al.}(2019)Gaita-Ari{\~n}o, Luis, Hill, and
  Coronado]{gaita2019molecular}
Gaita-Ari{\~n}o,~A.; Luis,~F.; Hill,~S.; Coronado,~E. Molecular spins for
  quantum computation. \emph{Nature chemistry} \textbf{2019}, \emph{11},
  301--309\relax
\mciteBstWouldAddEndPuncttrue
\mciteSetBstMidEndSepPunct{\mcitedefaultmidpunct}
{\mcitedefaultendpunct}{\mcitedefaultseppunct}\relax
\EndOfBibitem
\bibitem[Troiani and Affronte(2011)Troiani, and Affronte]{troiani2011molecular}
Troiani,~F.; Affronte,~M. Molecular spins for quantum information technologies.
  \emph{Chemical Society Reviews} \textbf{2011}, \emph{40}, 3119--3129\relax
\mciteBstWouldAddEndPuncttrue
\mciteSetBstMidEndSepPunct{\mcitedefaultmidpunct}
{\mcitedefaultendpunct}{\mcitedefaultseppunct}\relax
\EndOfBibitem
\bibitem[Stamp and Gaita-Arino(2009)Stamp, and Gaita-Arino]{stamp2009spin}
Stamp,~P.~C.; Gaita-Arino,~A. Spin-based quantum computers made by chemistry:
  hows and whys. \emph{Journal of Materials Chemistry} \textbf{2009},
  \emph{19}, 1718--1730\relax
\mciteBstWouldAddEndPuncttrue
\mciteSetBstMidEndSepPunct{\mcitedefaultmidpunct}
{\mcitedefaultendpunct}{\mcitedefaultseppunct}\relax
\EndOfBibitem
\bibitem[McAdams \latin{et~al.}(2017)McAdams, Ariciu, Kostopoulos, Walsh, and
  Tuna]{mcadams2017molecular}
McAdams,~S.~G.; Ariciu,~A.-M.; Kostopoulos,~A.~K.; Walsh,~J.~P.; Tuna,~F.
  Molecular single-ion magnets based on lanthanides and actinides: Design
  considerations and new advances in the context of quantum technologies.
  \emph{Coordination Chemistry Reviews} \textbf{2017}, \emph{346},
  216--239\relax
\mciteBstWouldAddEndPuncttrue
\mciteSetBstMidEndSepPunct{\mcitedefaultmidpunct}
{\mcitedefaultendpunct}{\mcitedefaultseppunct}\relax
\EndOfBibitem
\bibitem[Atzori \latin{et~al.}(2016)Atzori, Morra, Tesi, Albino, Chiesa,
  Sorace, and Sessoli]{atzori2016quantum}
Atzori,~M.; Morra,~E.; Tesi,~L.; Albino,~A.; Chiesa,~M.; Sorace,~L.;
  Sessoli,~R. Quantum coherence times enhancement in vanadium (IV)-based
  potential molecular qubits: the key role of the vanadyl moiety. \emph{Journal
  of the American Chemical Society} \textbf{2016}, \emph{138},
  11234--11244\relax
\mciteBstWouldAddEndPuncttrue
\mciteSetBstMidEndSepPunct{\mcitedefaultmidpunct}
{\mcitedefaultendpunct}{\mcitedefaultseppunct}\relax
\EndOfBibitem
\bibitem[Atzori \latin{et~al.}(2016)Atzori, Tesi, Morra, Chiesa, Sorace, and
  Sessoli]{atzori2016room}
Atzori,~M.; Tesi,~L.; Morra,~E.; Chiesa,~M.; Sorace,~L.; Sessoli,~R.
  Room-temperature quantum coherence and rabi oscillations in vanadyl
  phthalocyanine: toward multifunctional molecular spin qubits. \emph{Journal
  of the American Chemical Society} \textbf{2016}, \emph{138}, 2154--2157\relax
\mciteBstWouldAddEndPuncttrue
\mciteSetBstMidEndSepPunct{\mcitedefaultmidpunct}
{\mcitedefaultendpunct}{\mcitedefaultseppunct}\relax
\EndOfBibitem
\bibitem[Bader \latin{et~al.}(2014)Bader, Dengler, Lenz, Endeward, Jiang,
  Neugebauer, and Van~Slageren]{bader2014room}
Bader,~K.; Dengler,~D.; Lenz,~S.; Endeward,~B.; Jiang,~S.-D.; Neugebauer,~P.;
  Van~Slageren,~J. Room temperature quantum coherence in a potential molecular
  qubit. \emph{Nature Communications} \textbf{2014}, \emph{5}, 1--5\relax
\mciteBstWouldAddEndPuncttrue
\mciteSetBstMidEndSepPunct{\mcitedefaultmidpunct}
{\mcitedefaultendpunct}{\mcitedefaultseppunct}\relax
\EndOfBibitem
\bibitem[Graham \latin{et~al.}(2014)Graham, Zadrozny, Shiddiq, Anderson,
  Fataftah, Hill, and Freedman]{graham2014influence}
Graham,~M.~J.; Zadrozny,~J.~M.; Shiddiq,~M.; Anderson,~J.~S.; Fataftah,~M.~S.;
  Hill,~S.; Freedman,~D.~E. Influence of electronic spin and spin--orbit
  coupling on decoherence in mononuclear transition metal complexes.
  \emph{Journal of the American Chemical Society} \textbf{2014}, \emph{136},
  7623--7626\relax
\mciteBstWouldAddEndPuncttrue
\mciteSetBstMidEndSepPunct{\mcitedefaultmidpunct}
{\mcitedefaultendpunct}{\mcitedefaultseppunct}\relax
\EndOfBibitem
\bibitem[Atzori \latin{et~al.}(2018)Atzori, Benci, Morra, Tesi, Chiesa, Torre,
  Sorace, and Sessoli]{atzori2018structural}
Atzori,~M.; Benci,~S.; Morra,~E.; Tesi,~L.; Chiesa,~M.; Torre,~R.; Sorace,~L.;
  Sessoli,~R. Structural effects on the spin dynamics of potential molecular
  qubits. \emph{Inorganic Chemistry} \textbf{2018}, \emph{57}, 731--740\relax
\mciteBstWouldAddEndPuncttrue
\mciteSetBstMidEndSepPunct{\mcitedefaultmidpunct}
{\mcitedefaultendpunct}{\mcitedefaultseppunct}\relax
\EndOfBibitem
\bibitem[Bayliss \latin{et~al.}(2020)Bayliss, Laorenza, Mintun, Kovos,
  Freedman, and Awschalom]{bayliss2020optically}
Bayliss,~S.; Laorenza,~D.; Mintun,~P.; Kovos,~B.; Freedman,~D.~E.;
  Awschalom,~D. Optically addressable molecular spins for quantum information
  processing. \emph{Science} \textbf{2020}, \emph{370}, 1309--1312\relax
\mciteBstWouldAddEndPuncttrue
\mciteSetBstMidEndSepPunct{\mcitedefaultmidpunct}
{\mcitedefaultendpunct}{\mcitedefaultseppunct}\relax
\EndOfBibitem
\bibitem[Carretta \latin{et~al.}(2021)Carretta, Zueco, Chiesa,
  G{\'o}mez-Le{\'o}n, and Luis]{carretta2021perspective}
Carretta,~S.; Zueco,~D.; Chiesa,~A.; G{\'o}mez-Le{\'o}n,~{\'A}.; Luis,~F. A
  perspective on scaling up quantum computation with molecular spins.
  \emph{Applied Physics Letter} \textbf{2021}, \emph{118}, 240501\relax
\mciteBstWouldAddEndPuncttrue
\mciteSetBstMidEndSepPunct{\mcitedefaultmidpunct}
{\mcitedefaultendpunct}{\mcitedefaultseppunct}\relax
\EndOfBibitem
\bibitem[Li \latin{et~al.}(2021)Li, Xiong, Li, Jin, Zhang, Jiang, Ouyang, Wang,
  Wu, van Tol, \latin{et~al.} others]{li2021manipulation}
Li,~J.; Xiong,~S.-J.; Li,~C.; Jin,~B.; Zhang,~Y.-Q.; Jiang,~S.-D.;
  Ouyang,~Z.-W.; Wang,~Z.; Wu,~X.-L.; van Tol,~J., \latin{et~al.}  Manipulation
  of Molecular Qubits by Isotope Effect on Spin Dynamics. \emph{CCS Chemistry}
  \textbf{2021}, \emph{3}, 2548--2556\relax
\mciteBstWouldAddEndPuncttrue
\mciteSetBstMidEndSepPunct{\mcitedefaultmidpunct}
{\mcitedefaultendpunct}{\mcitedefaultseppunct}\relax
\EndOfBibitem
\bibitem[Nelson \latin{et~al.}(2020)Nelson, Zhang, Zhou, Rugg, Krzyaniak, and
  Wasielewski]{nelson2020cnot}
Nelson,~J.~N.; Zhang,~J.; Zhou,~J.; Rugg,~B.~K.; Krzyaniak,~M.~D.;
  Wasielewski,~M.~R. CNOT gate operation on a photogenerated molecular electron
  spin-qubit pair. \emph{Journal of Chemical Physics} \textbf{2020},
  \emph{152}, 014503\relax
\mciteBstWouldAddEndPuncttrue
\mciteSetBstMidEndSepPunct{\mcitedefaultmidpunct}
{\mcitedefaultendpunct}{\mcitedefaultseppunct}\relax
\EndOfBibitem
\bibitem[Thiele \latin{et~al.}(2014)Thiele, Balestro, Ballou, Klyatskaya,
  Ruben, and Wernsdorfer]{thiele2014electrically}
Thiele,~S.; Balestro,~F.; Ballou,~R.; Klyatskaya,~S.; Ruben,~M.;
  Wernsdorfer,~W. Electrically driven nuclear spin resonance in single-molecule
  magnets. \emph{Science} \textbf{2014}, \emph{344}, 1135--1138\relax
\mciteBstWouldAddEndPuncttrue
\mciteSetBstMidEndSepPunct{\mcitedefaultmidpunct}
{\mcitedefaultendpunct}{\mcitedefaultseppunct}\relax
\EndOfBibitem
\bibitem[Hussain \latin{et~al.}(2018)Hussain, Allodi, Chiesa, Garlatti, Mitcov,
  Konstantatos, Pedersen, De~Renzi, Piligkos, and
  Carretta]{hussain2018coherent}
Hussain,~R.; Allodi,~G.; Chiesa,~A.; Garlatti,~E.; Mitcov,~D.;
  Konstantatos,~A.; Pedersen,~K.~S.; De~Renzi,~R.; Piligkos,~S.; Carretta,~S.
  Coherent manipulation of a molecular Ln-based nuclear qudit coupled to an
  electron qubit. \emph{Journal of the American Chemical Society}
  \textbf{2018}, \emph{140}, 9814--9818\relax
\mciteBstWouldAddEndPuncttrue
\mciteSetBstMidEndSepPunct{\mcitedefaultmidpunct}
{\mcitedefaultendpunct}{\mcitedefaultseppunct}\relax
\EndOfBibitem
\bibitem[Hauser(2004)]{hauser2004ligand}
Hauser,~A. In \emph{Spin Crossover in Transition Metal Compounds I};
  G{\"u}tlich,~P., Goodwin,~H., Eds.; Springer Berlin Heidelberg: Berlin,
  Heidelberg, 2004; pp 49--58\relax
\mciteBstWouldAddEndPuncttrue
\mciteSetBstMidEndSepPunct{\mcitedefaultmidpunct}
{\mcitedefaultendpunct}{\mcitedefaultseppunct}\relax
\EndOfBibitem
\bibitem[Kepenekian \latin{et~al.}(2009)Kepenekian, Le~Guennic, and
  Robert]{kepenekian2009primary}
Kepenekian,~M.; Le~Guennic,~B.; Robert,~V. Primary Role of the Electrostatic
  Contributions in a Rational Growth of Hysteresis Loop in Spin-Crossover
  Fe(II) Complexes. \emph{Journal of the American Chemical Society}
  \textbf{2009}, \emph{131}, 11498--11502\relax
\mciteBstWouldAddEndPuncttrue
\mciteSetBstMidEndSepPunct{\mcitedefaultmidpunct}
{\mcitedefaultendpunct}{\mcitedefaultseppunct}\relax
\EndOfBibitem
\bibitem[Kepenekian \latin{et~al.}(2009)Kepenekian, Le~Guennic, and
  Robert]{kenepekian2009magnetic}
Kepenekian,~M.; Le~Guennic,~B.; Robert,~V. Magnetic bistability: From
  microscopic to macroscopic understandings of hysteretic behavior using ab
  initio calculations. \emph{Physical Review B} \textbf{2009}, \emph{79},
  094428\relax
\mciteBstWouldAddEndPuncttrue
\mciteSetBstMidEndSepPunct{\mcitedefaultmidpunct}
{\mcitedefaultendpunct}{\mcitedefaultseppunct}\relax
\EndOfBibitem
\bibitem[Tezgerevska \latin{et~al.}(2014)Tezgerevska, Alley, and
  Boskovic]{tezgerevska2014valence}
Tezgerevska,~T.; Alley,~K.~G.; Boskovic,~C. Valence tautomerism in metal
  complexes: Stimulated and reversible intramolecular electron transfer between
  metal centers and organic ligands. \emph{Coordination Chemistry Reviews}
  \textbf{2014}, \emph{268}, 23--40\relax
\mciteBstWouldAddEndPuncttrue
\mciteSetBstMidEndSepPunct{\mcitedefaultmidpunct}
{\mcitedefaultendpunct}{\mcitedefaultseppunct}\relax
\EndOfBibitem
\bibitem[Roseiro \latin{et~al.}(2022)Roseiro, Ben~Amor, and
  Robert]{roseiro2022combining}
Roseiro,~P.; Ben~Amor,~N.; Robert,~V. Combining Open-Shell Verdazyl Environment
  and Co(II) Spin-Crossover: Spinmerism in Cobalt Oxoverdazyl Compound.
  \emph{ChemPhysChem} \textbf{2022}, \emph{23}, e202100801\relax
\mciteBstWouldAddEndPuncttrue
\mciteSetBstMidEndSepPunct{\mcitedefaultmidpunct}
{\mcitedefaultendpunct}{\mcitedefaultseppunct}\relax
\EndOfBibitem
\bibitem[Roseiro \latin{et~al.}(2022)Roseiro, Petit, Robert, and
  Yalouz]{roseiro2022Qubit}
Roseiro,~P.; Petit,~L.; Robert,~V.; Yalouz,~S. Emergence of Spinmerism for
  Molecular Spin-Qubits Generation. \emph{ChemPhysChem} \textbf{2022},
  e202200478\relax
\mciteBstWouldAddEndPuncttrue
\mciteSetBstMidEndSepPunct{\mcitedefaultmidpunct}
{\mcitedefaultendpunct}{\mcitedefaultseppunct}\relax
\EndOfBibitem
\bibitem[Oms \latin{et~al.}(2010)Oms, Rota, Norel, Calzado, Rousselière,
  Train, and Robert]{oms2010beyond}
Oms,~O.; Rota,~J.-B.; Norel,~L.; Calzado,~C.~J.; Rousselière,~H.; Train,~C.;
  Robert,~V. Beyond Kahn's Model: Substituent and Heteroatom Influence on
  Exchange Interaction in a Metal-Verdazyl Complex. \emph{European Journal of
  Inorganic Chemistry} \textbf{2010}, \emph{2010}, 5373--5378\relax
\mciteBstWouldAddEndPuncttrue
\mciteSetBstMidEndSepPunct{\mcitedefaultmidpunct}
{\mcitedefaultendpunct}{\mcitedefaultseppunct}\relax
\EndOfBibitem
\bibitem[Brook \latin{et~al.}(2010)Brook, Richardson, Haller, Hundley, and
  Yee]{brook2010strong}
Brook,~D. J.~R.; Richardson,~C.~J.; Haller,~B.~C.; Hundley,~M.; Yee,~G.~T.
  Strong ferromagnetic metal–ligand exchange in a nickel
  bis(3{,}5-dipyridylverdazyl) complex. \emph{Chem. Commun.} \textbf{2010},
  \emph{46}, 6590--6592\relax
\mciteBstWouldAddEndPuncttrue
\mciteSetBstMidEndSepPunct{\mcitedefaultmidpunct}
{\mcitedefaultendpunct}{\mcitedefaultseppunct}\relax
\EndOfBibitem
\bibitem[Barclay \latin{et~al.}(2003)Barclay, Hicks, Lemaire, and
  Thompson]{barclay2003verdazyl}
Barclay,~T.~M.; Hicks,~R.~G.; Lemaire,~M.~T.; Thompson,~L.~K. Verdazyl Radicals
  as Oligopyridine Mimics:{\thinspace} Structures and Magnetic Properties of
  M(II) Complexes of 1,5-Dimethyl-3-(2,2`-bipyridin-6-yl)-6-oxoverdazyl (M =
  Mn, Ni, Cu, Zn). \emph{Inorganic Chemistry} \textbf{2003}, \emph{42},
  2261--2267\relax
\mciteBstWouldAddEndPuncttrue
\mciteSetBstMidEndSepPunct{\mcitedefaultmidpunct}
{\mcitedefaultendpunct}{\mcitedefaultseppunct}\relax
\EndOfBibitem
\bibitem[Rota \latin{et~al.}(2010)Rota, Le~Guennic, and Robert]{rota2010toward}
Rota,~J.-B.; Le~Guennic,~B.; Robert,~V. Toward Verdazyl Radical-Based
  Materials: Ab Initio Inspection of Potential Organic Candidates for
  Spin-Crossover Phenomenon. \emph{Inorganic Chemistry} \textbf{2010},
  \emph{49}, 1230--1237\relax
\mciteBstWouldAddEndPuncttrue
\mciteSetBstMidEndSepPunct{\mcitedefaultmidpunct}
{\mcitedefaultendpunct}{\mcitedefaultseppunct}\relax
\EndOfBibitem
\bibitem[Fleming \latin{et~al.}(2020)Fleming, Chung, Ponce, Brook, DaRos, Das,
  Ozarowski, and Stoian]{fleming2020valence}
Fleming,~C.; Chung,~D.; Ponce,~S.; Brook,~D. J.~R.; DaRos,~J.; Das,~R.;
  Ozarowski,~A.; Stoian,~S.~A. Valence tautomerism in a cobalt-verdazyl
  coordination compound. \emph{Chem. Commun.} \textbf{2020}, \emph{56},
  4400--4403\relax
\mciteBstWouldAddEndPuncttrue
\mciteSetBstMidEndSepPunct{\mcitedefaultmidpunct}
{\mcitedefaultendpunct}{\mcitedefaultseppunct}\relax
\EndOfBibitem
\bibitem[Brook \latin{et~al.}(2018)Brook, Fleming, Chung, Richardson, Ponce,
  Das, Srikanth, Heindl, and Noll]{brook2018anelectron}
Brook,~D. J.~R.; Fleming,~C.; Chung,~D.; Richardson,~C.; Ponce,~S.; Das,~R.;
  Srikanth,~H.; Heindl,~R.; Noll,~B.~C. An electron transfer driven magnetic
  switch: ferromagnetic exchange and spin delocalization in iron verdazyl
  complexes. \emph{Dalton Trans.} \textbf{2018}, \emph{47}, 6351--6360\relax
\mciteBstWouldAddEndPuncttrue
\mciteSetBstMidEndSepPunct{\mcitedefaultmidpunct}
{\mcitedefaultendpunct}{\mcitedefaultseppunct}\relax
\EndOfBibitem
\bibitem[Decurtins \latin{et~al.}(1984)Decurtins, Gütlich, Köhler, Spiering,
  and Hauser]{LIESST1984}
Decurtins,~S.; Gütlich,~P.; Köhler,~C.; Spiering,~H.; Hauser,~A.
  Light-induced excited spin state trapping in a transition-metal complex: The
  hexa-1-propyltetrazole-iron (II) tetrafluoroborate spin-crossover system.
  \emph{Chemical Physics Letters} \textbf{1984}, \emph{105}, 1--4\relax
\mciteBstWouldAddEndPuncttrue
\mciteSetBstMidEndSepPunct{\mcitedefaultmidpunct}
{\mcitedefaultendpunct}{\mcitedefaultseppunct}\relax
\EndOfBibitem
\bibitem[Maynau \latin{et~al.}(2002)Maynau, Evangelisti, Guihéry, Calzado, and
  Malrieu]{DOLOcode}
Maynau,~D.; Evangelisti,~S.; Guihéry,~N.; Calzado,~C.~J.; Malrieu,~J.-P.
  Direct generation of local orbitals for multireference treatment and
  subsequent uses for the calculation of the correlation energy. \emph{The
  Journal of Chemical Physics} \textbf{2002}, \emph{116}, 10060--10068\relax
\mciteBstWouldAddEndPuncttrue
\mciteSetBstMidEndSepPunct{\mcitedefaultmidpunct}
{\mcitedefaultendpunct}{\mcitedefaultseppunct}\relax
\EndOfBibitem
\bibitem[Aquilante \latin{et~al.}(2016)Aquilante, Autschbach, Carlson,
  Chibotaru, Delcey, De~Vico, Fdez.~Galván, Ferré, Frutos, Gagliardi,
  Garavelli, Giussani, Hoyer, Li~Manni, Lischka, Ma, Malmqvist, Müller, Nenov,
  Olivucci, Pedersen, Peng, Plasser, Pritchard, Reiher, Rivalta, Schapiro,
  Segarra-Martí, Stenrup, Truhlar, Ungur, Valentini, Vancoillie, Veryazov,
  Vysotskiy, Weingart, Zapata, and Lindh]{MOLCAS}
Aquilante,~F. \latin{et~al.}  Molcas 8: New capabilities for
  multiconfigurational quantum chemical calculations across the periodic table.
  \emph{Journal of Computational Chemistry} \textbf{2016}, \emph{37},
  506--541\relax
\mciteBstWouldAddEndPuncttrue
\mciteSetBstMidEndSepPunct{\mcitedefaultmidpunct}
{\mcitedefaultendpunct}{\mcitedefaultseppunct}\relax
\EndOfBibitem
\bibitem[Miralles \latin{et~al.}(1992)Miralles, Daudey, and Caballol]{DDCIref1}
Miralles,~J.; Daudey,~J.-P.; Caballol,~R. Variational calculation of small
  energy differences. The singlet-triplet gap in [Cu2Cl6]$^{2-}$.
  \emph{Chemical physics letters} \textbf{1992}, \emph{198}, 555--562\relax
\mciteBstWouldAddEndPuncttrue
\mciteSetBstMidEndSepPunct{\mcitedefaultmidpunct}
{\mcitedefaultendpunct}{\mcitedefaultseppunct}\relax
\EndOfBibitem
\bibitem[Miralles \latin{et~al.}(1993)Miralles, Castell, Caballol, and
  Malrieu]{DDCIref2}
Miralles,~J.; Castell,~O.; Caballol,~R.; Malrieu,~J.-P. Specific CI calculation
  of energy differences: Transition energies and bond energies. \emph{Chemical
  physics} \textbf{1993}, \emph{172}, 33--43\relax
\mciteBstWouldAddEndPuncttrue
\mciteSetBstMidEndSepPunct{\mcitedefaultmidpunct}
{\mcitedefaultendpunct}{\mcitedefaultseppunct}\relax
\EndOfBibitem
\bibitem[{Ben Amor} and Maynau(1998){Ben Amor}, and Maynau]{CASDIcode}
{Ben Amor},~N.; Maynau,~D. Size-consistent self-consistent configuration
  interaction from a complete active space. \emph{Chemical Physics Letters}
  \textbf{1998}, \emph{286}, 211--220\relax
\mciteBstWouldAddEndPuncttrue
\mciteSetBstMidEndSepPunct{\mcitedefaultmidpunct}
{\mcitedefaultendpunct}{\mcitedefaultseppunct}\relax
\EndOfBibitem
\bibitem[Yalouz \latin{et~al.}(2022)Yalouz, Gullin, and
  Sekaran]{yalouz2022quantnbody}
Yalouz,~S.; Gullin,~M.~R.; Sekaran,~S. QuantNBody: a Python package for quantum
  chemistry and physics to build and manipulate many-body operators and wave
  functions. \emph{Journal of Open Source Software} \textbf{2022}, \emph{7},
  4759\relax
\mciteBstWouldAddEndPuncttrue
\mciteSetBstMidEndSepPunct{\mcitedefaultmidpunct}
{\mcitedefaultendpunct}{\mcitedefaultseppunct}\relax
\EndOfBibitem
\bibitem[Ghosh \latin{et~al.}(2003)Ghosh, Bill, Weyherm{\"u}ller, Neese, and
  Wieghardt]{ghosh2003noninnocence}
Ghosh,~P.; Bill,~E.; Weyherm{\"u}ller,~T.; Neese,~F.; Wieghardt,~K.
  Noninnocence of the Ligand Glyoxal-bis (2-mercaptoanil). The Electronic
  Structures of [Fe (gma)] 2,[Fe (gma)(py)] py,[Fe (gma)(CN)] 1-/0,[Fe (gma)
  I], and [Fe (gma)(PR3) n](n= 1, 2). Experimental and Theoretical Evidence for
  “Excited State” Coordination. \emph{Journal of the American Chemical
  Society} \textbf{2003}, \emph{125}, 1293--1308\relax
\mciteBstWouldAddEndPuncttrue
\mciteSetBstMidEndSepPunct{\mcitedefaultmidpunct}
{\mcitedefaultendpunct}{\mcitedefaultseppunct}\relax
\EndOfBibitem
\bibitem[Tanabe and Sugano(1954)Tanabe, and Sugano]{tanabe1954onthe}
Tanabe,~Y.; Sugano,~S. On the Absorption Spectra of Complex Ions II.
  \emph{Journal of the Physical Society of Japan} \textbf{1954}, \emph{9},
  766--779\relax
\mciteBstWouldAddEndPuncttrue
\mciteSetBstMidEndSepPunct{\mcitedefaultmidpunct}
{\mcitedefaultendpunct}{\mcitedefaultseppunct}\relax
\EndOfBibitem
\bibitem[Delgado \latin{et~al.}(2018)Delgado, Tissot, Guénée, Hauser,
  Valverde-Muñoz, Seredyuk, Real, Pillet, Bendeif, and
  Besnard]{delgado2018verylong}
Delgado,~T.; Tissot,~A.; Guénée,~L.; Hauser,~A.; Valverde-Muñoz,~F.~J.;
  Seredyuk,~M.; Real,~J.~A.; Pillet,~S.; Bendeif,~E.-E.; Besnard,~C. Very
  Long-Lived Photogenerated High-Spin Phase of a Multistable Spin-Crossover
  Molecular Material. \emph{Journal of the American Chemical Society}
  \textbf{2018}, \emph{140}, 12870--12876\relax
\mciteBstWouldAddEndPuncttrue
\mciteSetBstMidEndSepPunct{\mcitedefaultmidpunct}
{\mcitedefaultendpunct}{\mcitedefaultseppunct}\relax
\EndOfBibitem
\bibitem[Sousa \latin{et~al.}(2013)Sousa, de~Graaf, Rudavskyi, Broer, Tatchen,
  Etinski, and Marian]{sousa2013ultrafast}
Sousa,~C.; de~Graaf,~C.; Rudavskyi,~A.; Broer,~R.; Tatchen,~J.; Etinski,~M.;
  Marian,~C.~M. Ultrafast Deactivation Mechanism of the Excited Singlet in the
  Light-Induced Spin Crossover of [Fe (2, 2'-bipyridine) 3] 2+.
  \emph{Chemistry--A European Journal} \textbf{2013}, \emph{19},
  17541--17551\relax
\mciteBstWouldAddEndPuncttrue
\mciteSetBstMidEndSepPunct{\mcitedefaultmidpunct}
{\mcitedefaultendpunct}{\mcitedefaultseppunct}\relax
\EndOfBibitem
\bibitem[Alías-Rodríguez \latin{et~al.}(2022)Alías-Rodríguez,
  Huix-Rotllant, and de~Graaf]{alias2022quantum}
Alías-Rodríguez,~M.; Huix-Rotllant,~M.; de~Graaf,~C. Quantum dynamics
  simulations of the thermal and light-induced high-spin to low-spin relaxation
  in Fe(bpy)3 and Fe(mtz)6. \emph{Faraday Discuss.} \textbf{2022}, \emph{237},
  93--107\relax
\mciteBstWouldAddEndPuncttrue
\mciteSetBstMidEndSepPunct{\mcitedefaultmidpunct}
{\mcitedefaultendpunct}{\mcitedefaultseppunct}\relax
\EndOfBibitem
\bibitem[Messaoudi \latin{et~al.}(2006)Messaoudi, Robert, Guihéry, and
  Maynau]{messaoudi2006correlated}
Messaoudi,~S.; Robert,~V.; Guihéry,~N.; Maynau,~D. Correlated ab Initio Study
  of the Excited State of the Iron-Coordinated-Mode Noninnocent
  Glyoxalbis(mercaptoanil) Ligand. \emph{Inorganic Chemistry} \textbf{2006},
  \emph{45}, 3212--3216, PMID: 16602777\relax
\mciteBstWouldAddEndPuncttrue
\mciteSetBstMidEndSepPunct{\mcitedefaultmidpunct}
{\mcitedefaultendpunct}{\mcitedefaultseppunct}\relax
\EndOfBibitem
\bibitem[Guihéry \latin{et~al.}(2008)Guihéry, Robert, and
  Neese]{guihery2008abinitio}
Guihéry,~N.; Robert,~V.; Neese,~F. Ab Initio Study of Intriguing Coordination
  Complexes: A Metal Field Theory Picture. \emph{The Journal of Physical
  Chemistry A} \textbf{2008}, \emph{112}, 12975--12979, PMID: 18811129\relax
\mciteBstWouldAddEndPuncttrue
\mciteSetBstMidEndSepPunct{\mcitedefaultmidpunct}
{\mcitedefaultendpunct}{\mcitedefaultseppunct}\relax
\EndOfBibitem
\end{mcitethebibliography}

\end{document}